# A dynamical approach to generate chaos in a micromechanical resonator

Martial Defoort*, Libor Rufer, Laurent Fesquet and Skandar Basrour*
Univ. Grenoble Alpes, CNRS, Grenoble INP, TIMA, 38000 Grenoble, France

**Chaotic systems, presenting complex and non-reproducible dynamics, may be found in nature from the interaction between planets to the evolution of the weather, but can also be tailored using current technologies for advanced signal processing. However, the realization of chaotic signal generators remains challenging, due to the involved dynamics of the underlying physics. In this paper, we experimentally and numerically present a disruptive approach to generate a chaotic signal from a micromechanical resonator. This technique overcomes the long-established complexity of controlling the buckling in micro/nano-mechanical structures by modulating either the amplitude or the frequency of the driving force applied to the resonator in the nonlinear regime. The experimental characteristic parameters of the chaotic regime, namely the Poincaré sections and Lyapunov exponents, are directly comparable to simulations for different configurations. These results confirm that this dynamical approach is transposable to any kind of micro/nano-mechanical resonators, from accelerometers to microphones. We demonstrate a direct application exploiting the mixing properties of the chaotic regime by transforming an off-the-shelf microdiaphragm into a true random number generator conformed to the National Institute of Standards and Technology specifications. The versatility of this original method opens new paths to combine chaos' unique properties with microstructures' exceptional sensitivity leading to emergent microsystems.**

## Introduction

Micro and Nano ElectroMechanical Systems (M/NEMS) have become essential building blocks for the development of modern technologies due to their small size, low cost and compatibility with microelectronics, with various applications such as sensors[1], actuators[2] or clocks[3]. Beside their exceptional properties optimized for high-end products, these mechanical structures are also remarkable tools for fundamental physics, both for classical[4] or quantum investigations[5]. This duality triggered studies aiming to exploit singular properties of nonlinear dynamics for direct applications in micro/nano-structures, by taking advantage of the synchronization phenomenon to enhance MEMS accelerometers[6], using internal resonances to reduce frequency drifts for timing purposes[7] or operating microcantilevers in the nonlinear regime for mass sensing applications[8].

Amongst these various nonlinear phenomena improving M/NEMS performance, the chaotic regime features some of the most singular properties, addressing the complex needs of true random number generators[9], secured communications[10] or sensing applications[11,12], but has yet to be implemented with a convenient, generic approach. Chaotic behaviors describe a large variety of involved interactions, characterizing the evolution of cosmic entities[13] or interpreting the unpredictability of the weather[14]. Inherently complex, chaotic signals share some properties with noise, having a broad frequency spectrum and an apparent irreproducibility due to its exponentially sensitivity to the initial conditions. This unique property generated various works in electrical circuits, which were amongst the first physical chaotic systems tailored[15], and on the generation of chaos in lasers for optical telecommunications applications[16].

Due to their versatility and their large, tunable nonlinearity[17], M/NEMS are prime candidates for chaos studies and applications. The Duffing nonlinearity, defined as a cubic stiffness, is the most common source of chaos in M/NEMS, which is achieved by buckling the mechanical structure using specific geometries[18], materials[19] or configurations[20]. The system enters in a bistable configuration, between the buckled up and down states, defined by a double-well potential. By driving such device with a large enough force, the structure may switch between the two states, and using the appropriate driving frequency the system experiences a chaotic regime. However, buckling micro/nano-structures at will is demanding, and while some buckled MEMS devices demonstrated experimentally a chaotic regime[21–24], research on the topic is mostly performed only analytically or numerically[25–27]. Among the experimental issues, the realization of the buckling states usually requires high voltages of tens to hundreds of Volts[21,22,28] and the large amplitudes involved outrange the linear regime of commonly used transduction schemes[25]. Besides, the buckled property of the structure itself greatly reduces the range of applications of the generated chaos, and only a few works suggest non-buckled alternatives[24,29], remaining difficult to accomplish or to investigate.

In this paper, we demonstrate experimentally the realization of a chaotic system based on the modulation of the driving signal[30] and present a direct application. The only two requirements are 1) obtaining the Duffing regime, present in most micro/nano-resonators for a large enough drive, 2) performing either Amplitude Modulation (AM) or Frequency Modulation (FM) on the driving force. This technique is readily applicable in most current devices, as neither fabrication process nor specific geometries are required. Considering the latest micro/nano-technological advances, for which each application results in optimized designs, this implementation of chaos opens a path to combine the intrinsic sensitivity of micro/nano-devices with the various properties of chaos, leading to new, emergent systems.

In the following paragraphs, we present the working principle of this chaotic regime. We characterize its properties through the bifurcation diagrams, the Poincaré sections and the Lyapunov exponents mapping the range of the chaotic regime, in AM and FM configurations. We compare these results with simulations using no adjusting parameters. Finally, we demonstrate that this system may be used as a true random number generator conformed to the National Institute of Standards and Technology (NIST) specifications.

## Results
### Working principle and device

The generation of a chaotic regime is mostly performed with multi-stable systems. In the mechanical domain, this property is found in buckled structures, which are built by applying a constraint to the device, leading to a *static* bistability (top panel of Fig.1 a). These states are defined by a static double-well potential, each well describing the buckled up and down states[21]. A chaotic regime emerges from the complex evolution between the buckled states due to an additional driving force.

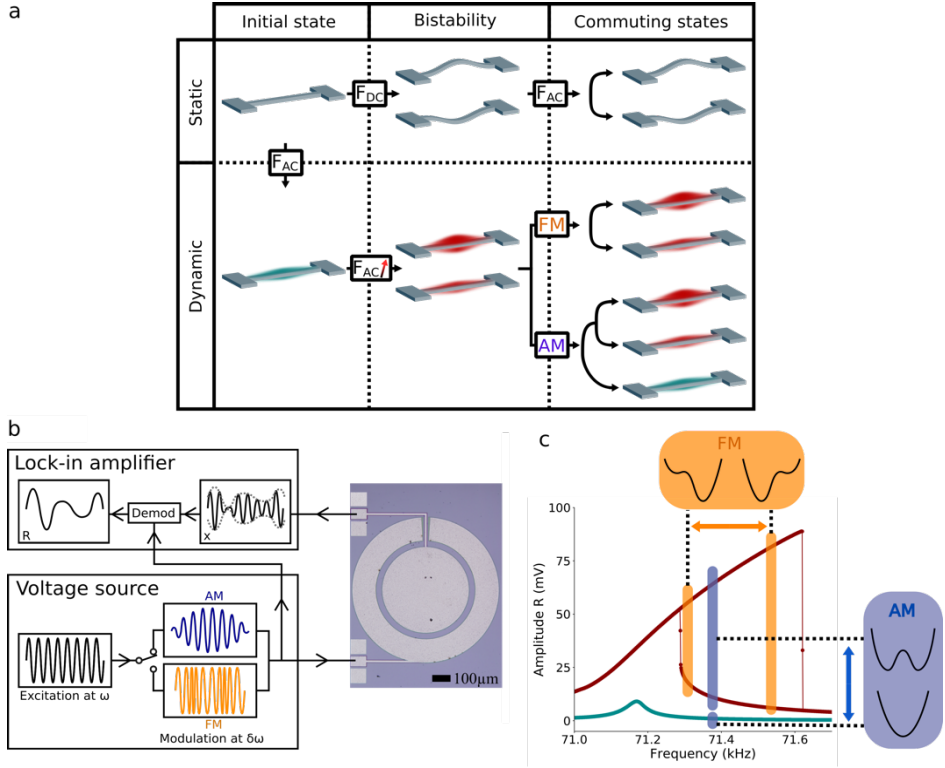

**Figure 1: The dynamical bistability for the generation of a chaotic regime in a nonlinear resonator. a** Comparison between static and dynamical bistability for chaos generation in a nonlinear resonator, here represented by a doubly-clamped beam. In the static case, a resonator reaches bistability through buckling by applying a static force $F_{DC}$, while in the dynamical case the alternative driving force $F_{AC}$ is increased to enter in the Duffing regime. By modulating the driving signal, the system evolves between the different states available, which results in a chaotic regime for an appropriate set of parameters. **b** Experimental setup. The MEMS (top view photography) is driven by a voltage source which provides AM or FM configurations. The LIA demodulates the displacement x of the MEMS at the frequency of the source output to retrieve the amplitude R. **c** Amplitude response of the MEMS in the linear and Duffing regime (respectively dark cyan and dark red at 10 mV and 100 mV drive). The schematic illustrates qualitatively the evolution of the potentials involved for both amplitude and frequency modulations (AM and FM).

However, most of these structures may instead be driven close to the resonance with a strong enough force to reveal their nonlinear behavior with the Duffing regime. In this regime, the resonator vibrates either with a large or a low amplitude[17] (bottom panel of Fig. 1 a), leading to a *dynamical* bistability described by a dynamical double-well potential. As for the buckled structure, the Duffing regime may be used as the starting point to generate chaos, by modulating the driving signal to commute between the different available states[30]. But in contrast with the static case, this dynamical bistability only rely on intrinsic properties present in most M/NEMS, making this chaotic regime achievable for off-the-shelf devices. In this paper, we present the case of a typical circular diaphragm.

The proof-of-concept structure was fabricated using a standard multi-user MEMS process provided by the company Memscap under the brand name PiezoMUMPs. This process is CMOS compatible and allows multilayered structures of dimensions complying with our specifications[31]. The MEMS is a silicon-on-insulator based circular diaphragm, with a radius of 400 µm and a thickness of 10 µm (Fig. 1 b and Supplementary Information (SI), Fig. S1), placed under vacuum at room temperature. The first flexural mode of the structure is actuated with a voltage source and detected with Lock-In Amplifier (LIA) assisted by a current amplifier, using the outer and inner electrodes coupled to the resonator, comprised of a piezoelectric 500 nm thick AlN layer. At first order this resonator is described by a 1D model for which the dynamics corresponds to the canonical equation:

$$\ddot{x} + \Delta\omega\,\dot{x} + \omega_0^2\,x + \alpha\,x^3 = \frac{F}{m}\cos(\omega\,t) \quad (1)$$

with $\Delta\omega = 2\pi\,\Delta f$, $\omega_0 = 2\pi\,f_0$, $\alpha_n = \frac{3\,\alpha}{8\,\omega_0}$ and $m$ being respectively the bandwidth, the angular natural resonance frequency, the Duffing nonlinear coefficient and the mass of the resonator, with $Q = \frac{f}{\Delta f}$ its quality factor. With the present device, we fit a natural resonance frequency $f_0 = 71.2$ kHz, a bandwidth $\Delta f = 50$ Hz leading to a quality factor $Q = 1420$, and a nonlinear coefficient $\alpha_n = 54$ kHz/V$^2$ (see SI, Fig. S2). The structure is driven at an angular frequency $\omega = 2\pi\,f$ by a force $F$. In our case, this force results of a voltage difference applied on the diaphragm's piezoelectric layers. Note that the chaotic regime presented in this paper does not depend on the transduction mechanism and remains compatible with capacitive or optical techniques. To characterize the vibrating signal of the mechanical structure, we perform a rotating frame approximation at the driving frequency (see SI, note 1), giving at first order:

$$\dot{R} = -\frac{\Delta\omega}{2}R - \frac{F}{2\,m\,\omega}\sin\varphi \quad (2.a)$$

$$\dot{\varphi} = \omega_0 - \omega + \alpha_n R^2 - \frac{F}{2\,m\,\omega\,R}\cos\varphi \quad (2.b)$$

with $R$ the demodulated amplitude - the envelope of the signal, and $\varphi$ its phase delay with the driving force. The nonlinear term $\alpha_n R^2$ shifts the resonance frequency which results in a hysteresis, at the essence of the Duffing nonlinearity (Fig. 1 c), where the MEMS evolves either with a large or a low amplitude for the same driving frequency. The state of the resonator in the hysteresis depends on the history of the system, and the amplitude of vibration may switch between the high and the low level in the presence of a perturbation[32,33]. At this point, (2) only describes a bistable potential: an additional parameter is necessary to create an evolution of this dynamical system and possibly generate chaos.

The first configuration we describe performs an amplitude modulation on the driving force, namely $F \rightarrow F\frac{1+\cos\delta\omega t}{2}$, with $\delta\omega = 2\pi\,\delta f$ the angular modulation frequency of the signal, directly transposable to (2) (see SI, note 1). The modulation having a 100% depth, the force experienced by the system oscillates from 0 to $F$ at the rate $\delta\omega$. If $F$ is large enough to open the hysteresis of the Duffing resonator, the system is oscillating between a linear regime with a single-well potential to a nonlinear regime with a double-well potential (Fig. 1, c, blue areas).

The second configuration consists in a frequency modulation where the driving phase becomes $\omega t \rightarrow \omega t + \sin(\delta\omega t)$, which has a modulation index of 1. In (2) it follows $\omega \rightarrow \omega + \delta\omega \cos(\delta\omega t)$ (see SI, note 1). This essentially corresponds to back and forth sweeps through the resonance, such that the system may switch between the high and low amplitude states in the Duffing regime (Fig. 1, c, orange areas).

For slow modulations ($\delta\omega \ll \Delta\omega$), the system may switch between its two states but with no additional exotic behavior. However, when the modulation rate becomes comparable to the inverse of the system's time response, the resonator's dynamics becomes more complex and new physics may emerge. In the following section we consider a modulation rate of three times the resonator's bandwidth.

**Bifurcation diagram and Poincaré sections**

In order to characterize the evolution of a system from the periodic to the chaotic regime, a common approach is to perform a bifurcation diagram[34]. It consists in a stroboscopic view of the amplitude of a signal, sliced at the modulation frequency $\delta f$, as a given parameter is swept (in our case, the driving frequency $f$). These bifurcation diagrams enable to characterize the route to chaos of the system, which usually consists in an increasing number of harmonics in the signal, until reaching the chaotic regime with a broad frequency spectrum.

We performed a bifurcation diagram as a function of the driving frequency close to the natural resonance frequency (Fig. 2). At first, the modulation of the driving signal is accurately reproduced by the structure, with a low distortion (Fig. 2, a). However, as the driving frequency progresses through the hysteresis, the Duffing nonlinearity alters the mechanical response to the modulated driving signal, leading to a more complex yet periodic amplitude of vibration with higher harmonics (Fig. 2, b-c) until reaching a chaotic regime (Fig. 2, d). A closer look at the bifurcation diagram before entering in the chaotic regime (Fig. 2, f) reveals a period-doubling route to chaos, where the subharmonics present in the signal increase by a factor of two at each bifurcation point. In the AM configuration we measured period-doubling bifurcations up to $\frac{\delta f}{8}$. This route to chaos displays universal properties such as the constants of Feigenbaum[35], ultimately enabling to predict the threshold value after which the chaotic regime appears.

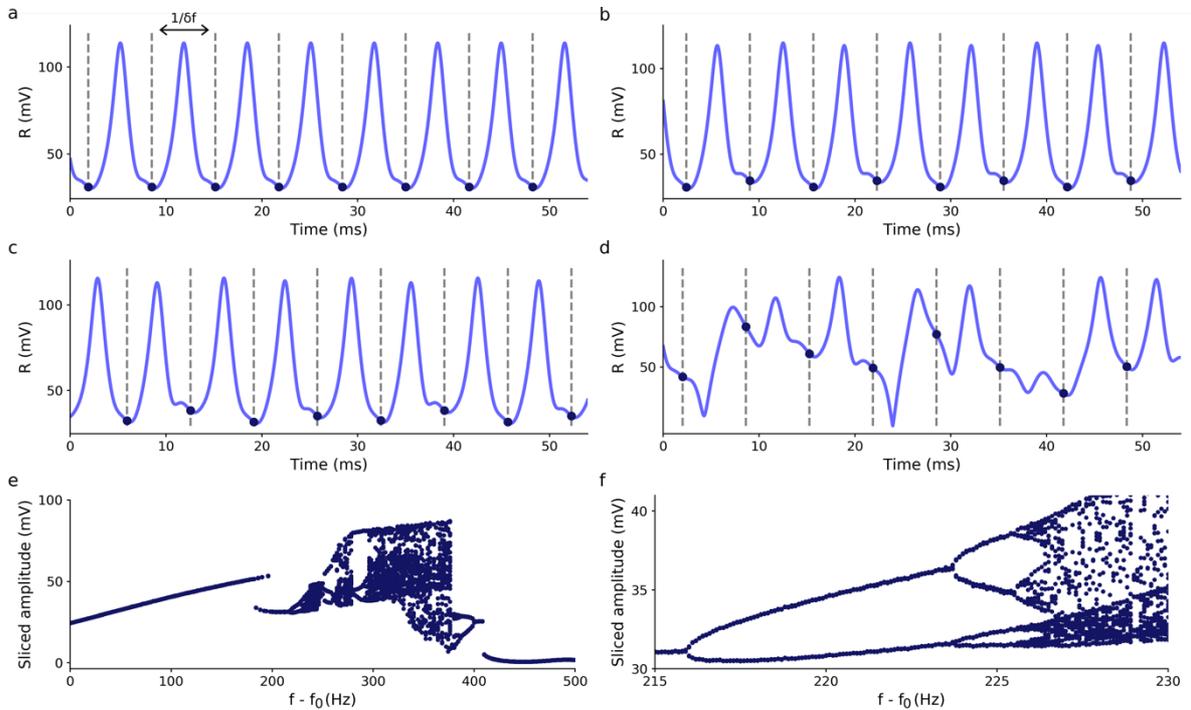

**Figure 2: Experimental bifurcation diagram in the AM configuration for a driving signal of 0.5 V and a modulation of 151 Hz.** To characterize the route to chaos, experimental data are sliced (dark blue dots) with a period of $1/\delta f$ (dashed lines) for different driving frequencies. Typical mechanical responses are presented (from **a** to **d**) for detuning frequencies $f - f_0$ of respectively 215 ; 220 ; 225 ; 340 Hz, which highlights the presence of 1, 2 and 4 harmonics in the signal, until reaching the chaotic regime. The projection of the sliced data for each driving frequency forms a bifurcation diagram (**e**). Looking closer at its evolution before entering in chaos (**f**), we observe the period-doubling route to chaos with period-doubling bifurcations up to $\frac{\delta f}{8}$.

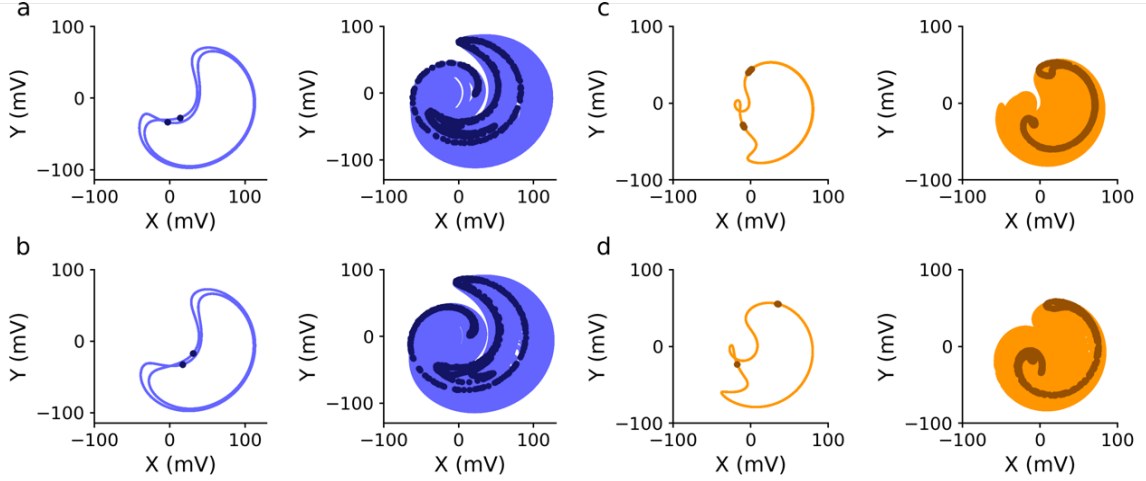

**Figure 3: Phase space and Poincaré sections of the linear and chaotic response of the resonator.** The experimental phase spaces of the AM and FM configurations (**a** and **c**) are compared with numerical results (respectively **b** and **d**). In the linear regime (left section of each panel), the periodic signal is in a limit cycle configuration and repeat itself endlessly (light color lines). The stroboscopic analysis reveals two periods in this example (dark color points). In the chaotic regime (right section of each panel), the signal is no longer periodic and fills up the phase space. The Poincaré section is composed of an infinite number of non-overlapping points which exhibit a specific signature. This signature is reproducible even in the chaotic regime, as demonstrated by the numerical analysis performed with no adjustable parameters. These signals were generated with a modulation rate of 151 Hz with different (drive, frequency detuning) sets of parameters. In the AM configuration we used (0.5 V, 220 Hz) in the periodic regime and (0.5 V, 340 Hz) in the chaotic regime. In the FM case we used (0.15 V, 155 Hz) in the periodic regime and (0.15 V, 175 Hz) in the chaotic regime.

This stroboscopic view is also at the basis of the Poincaré sections, which extract order from the apparent noisy structure of a chaotic signal[36]. Just like bifurcation diagrams, a long enough dataset is sliced at the modulation frequency with an arbitrary initial time. The sliced data are gathered on a graph presenting the phase space of the system, commonly shown as displacement versus velocity axis. In the present scenario, the variables are the in-phase ($X = R \cos \varphi$) and the out-of-phase ($Y = R \sin \varphi$) components of (2). For a given combination of fixed parameters, the global shape of a Poincaré section has a specific, reproducible signature. For a periodic signal, it represents a finite set of overlapping points, depending on the periodicity. For a chaotic signal, it forms a pattern of non-overlapping points (Fig. 3). The longer the dataset, the more precise the pattern of the chaotic Poincaré section becomes. Exploiting (2) with either amplitude or frequency modulation, the Poincaré sections of the chaotic signals were directly simulated with the measured experimental parameters of the system, reproducing quantitatively the experimental data.

While raw chaotic signals appear as random, the specific signatures of the Poincaré section analysis leave clues, partially revealing the nature of the chaotic system. Being able to reproduce numerically their shapes gives a lever to build a more complex chaos with potential applications in cryptography.

**Lyapunov exponents and TRNG**

One of the main aspects of a chaotic signal is its unpredictability and irreproducibility. These properties are related to the system's sensitivity to initial conditions, often categorized within three scenarios. In a damped oscillator, any initial mismatch between two similar measurements will progressively shrink over time. In a driven resonator, an initial phase delay mismatch will remain constant over time. In a chaotic resonator, any mismatch will increase over time. This property is characterized by the Lyapunov exponent $\lambda$, following $\delta z(t) = \delta z_0 \, e^{\lambda t}$ with $\delta z$ the distance between two initially close trajectories[37], resulting in: 1) converging trajectories ($\lambda < 0$), 2) conservative trajectories ($\lambda = 0$), 3) diverging trajectories ($\lambda > 0$). The measurement of this exponent is based on finding neighbor states in an acquired signal. In a periodic signal, any arbitrary picked state is reproduced after one period. Hence, there is one new neighbor state after each period. For a chaotic signal, a long enough acquisition ensures to eventually find close states, giving access to the local maximum Lyapunov exponent (SI, Fig. S3). After an even longer measurement, enough pairs of different initial states are gathered, enabling to compute the global Lyapunov exponent[38].

By changing the driving force and frequency for a fixed modulation rate, we mapped the regions where the MEMS has a chaotic response through the measurement of this Lyapunov exponent both experimentally and numerically (Fig. 4). We chose three modulation rates corresponding to one, two and three times the bandwidth of the linear resonator, for which the associated maps present different trends depending on the modulation configuration. In the AM case, the chaotic region appears to be bondless, and while the presented results stop for an equivalent force of 2 V, we kept measuring chaotic signals with the same modulation rates up to 3.5 V. In the FM configuration, the chaotic regime requires a lower force for a similar modulation rate. The chaotic region gets also broader with the modulation rate, but is confined in force and we could not measure any chaos above the voltages presented. This difference in the force range between AM and FM can be understood by looking back at the schematics of Fig. 1, c. In the AM configuration, for each driving force there will always be a detuning frequency within the hysteresis for which the double well potential will present a close-to-symmetric shape, enabling to access the two wells. In the FM configuration, we restricted our study to the case of a modulation index of 1. Therefore, the frequency span of the modulation is fixed at a given modulation rate while the size of the hysteresis increases with the driving force. The distance between the two wells of the hysteresis becomes then out of reach.

In both cases, the averaged Lyapunov exponent grows with the modulation rate, which indicates a smaller memory time. The experimental chaotic regions defined by $\lambda$ are quantitatively reproduced numerically, demonstrating that the simple system (2) is precise enough to encompass the behavior of the chaotic resonator despite the great variation between each modulation rate. Since no assumptions was made regarding the geometry or nature of the resonator, it is readily applicable to any Duffing resonator - right and top axis of Fig. 4 were normalized to ease it (see SI, note 2). The modulation rate to obtain chaos is not limited to the presented results, this regime is also achievable for larger drives with a modulation rate of up to a least 30 times the bandwidth (see SI, Fig. S4).

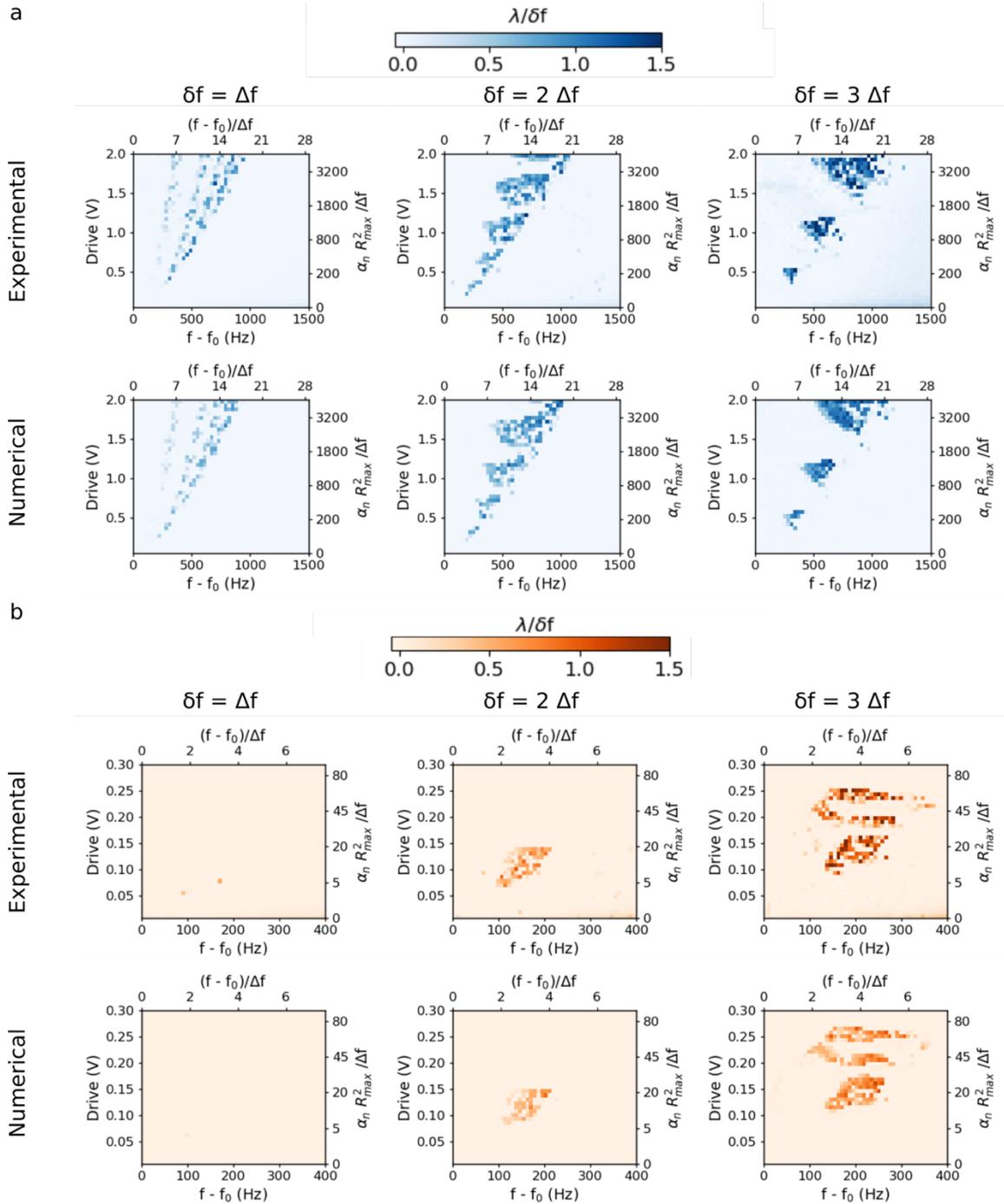

**Figure 4: Mapping of the Lyapunov exponents in both AM and FM configurations.** A positive global Lyapunov exponent is characteristic of a chaotic regime, and we use its value to determine the chaotic range as a function of the driving frequency and the driving force for modulation rates of one, two and three times the bandwidth of the system, in both AM (**a**) and FM (**b**) configurations. Note that the Lyapunov exponents are normalized to the modulation rate and right/top axis are normalized to the bandwidth of the system for a more comprehensive view of the phenomenon. The experimental Lyapunov maps are compared with numerical simulations using the measured experimental parameters.

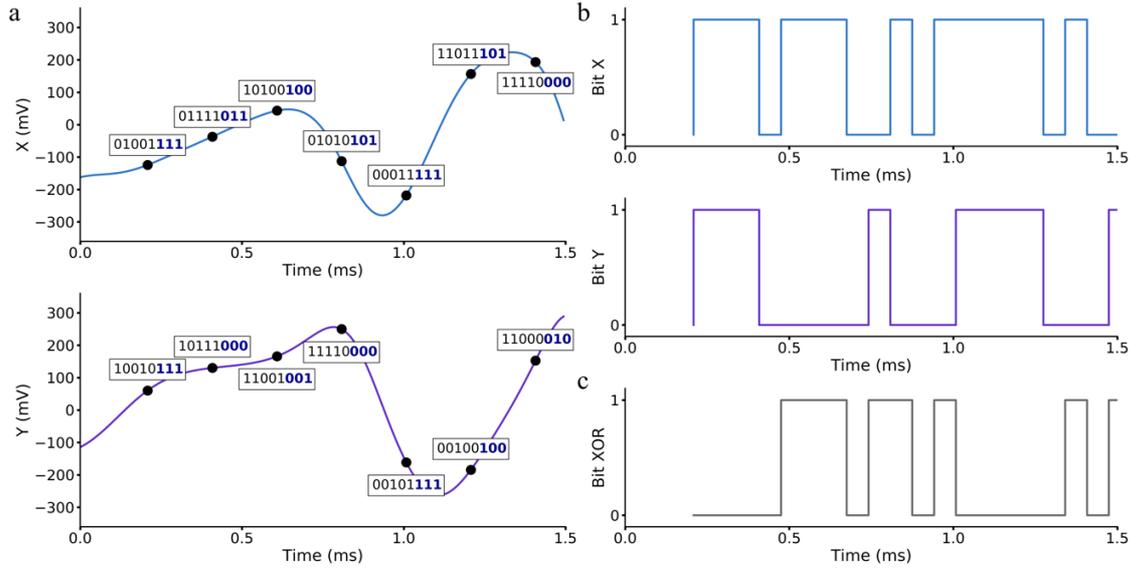

**Figure 5: Signal processing for the TRNG application.** Slicing the data with a sampling rate of 5 kHz, we convert the analog signal of both X and Y using a virtual 8-bits ADC (**a**). We then extract the three least significant bits of each measurement (bold, blue) and generate a bit sequence (**b**). We finally combine both X and Y sequences through a XOR function to generate the final bit stream, producing a random sequence at a rate of 15 kb/s (**c**). The chaotic signal was obtained for 3.5 V drive, 1250 Hz frequency detuning in the AM configuration at a modulation rate of 1.5 kHz.

The Lyapunov exponent describes the memory of the system. As such, the prediction of the evolution of a chaotic signal is limited by the measurement precision. This feature is well suited for true random number generation, and physical chaotic systems have already proven to pass standard randomness tests[9,39–41] such as NIST SP 800-22[42]. By itself, chaos only provides a complex way for mixing an initial state, which is deterministic, predictable and is the source of some numerical Pseudo-Random Number Generators[43] (PRNG). However, if the initial state is noisy, the predictability exponentially vanishes in the chaotic regime. Therefore, the combination of a stochastic seed (e.g. intrinsic thermomechanical noise or frequency fluctuations[44]) with the exponential sensitivity of chaos may turn the system into a True Random Number Generators (TRNG). To demonstrate this application, we performed the NIST tests on the output signal of our MEMS in the chaotic regime, which is converted into a binary form through a numerical Analog to Digital Convertor (ADC). To add up with the mixing properties of the chaos, we used a common process consisting in selecting the Least Significant Bits (LSB) of an 8-bits ADC digitizing the analog chaotic signal[39,40] (Fig. 5, a). The sampling rate can then be much faster than $\lambda$. In our case, we sampled at 5 kHz both X and Y components of the signal (Fig. 5, b). Keeping solely the 3-LSB of each measurement, we then performed a XOR function between X and Y sequences, known to improve the randomness of a bit stream (Fig. 5, c).

We finally obtained a sequence of 75 Mb divided into 75 sequences of 1 MB, processed through the NIST tests, showing that the chaotic system passes all the tests (see SI, Table 1) and delivering a random bit stream at a rate of 15 kb/s. As the chaotic system seed is noisy, a stochastic model of the entropy extraction can be obtained in order to prove the true system randomness. The performances of this TRNG can be tuned according to different needs, such as bit rate or power consumption. Because the method to generate chaos is non-invasive, most current M/NEMS devices initially designed for specific goals such as accelerometers or gyroscopes could additionally be used for true random number generation.

We stress on the fact that the MEMS structure used in this paper was not designed nor optimized for this chaotic regime. An improvement in chaos complexity, rate and power consumption may easily be performed through a higher Duffing nonlinearity, a higher resonance frequency and a lower bandwidth, while maintaining a qualitative understanding of the system.

In parallel to our work, a recent study also highlighted the potential of the dynamical bistability for chaos generation through a two drive tones[45]. However, they used a different approach and their results only focused on a specific working regime, with a distinct interpretation on the occurrence of chaos. In contrast, our alternative method allows us to draw a comprehensive overview of the phenomena at stake by illuminating the role of the different parameters of the system. In particular, our results point toward new opportunities and implementations impacting future technologies, which we illustrate with the generation of true random numbers, pillar of modern security and data protection.

## Discussion

Chaotic systems are known to present unique properties for cryptographic applications, which we illustrate in this paper with an experimental demonstration of an original chaotic MEMS based TRNG. This affinity between chaos and secured communication reaches a capstone with chaos synchronization[16], at the core of most research on chaotic lasers for telecommunication applications. While the optical domain benefits from a large bandwidth for unequalled data rate, micro/nano-mechanical resonators are tunable over orders of magnitude with a resolution below the ppm, providing a large number of cryptographic keys, essential for secured transmissions. Beside cryptography, the capacity to build a chaotic system from current mechanical structures opens new perspectives to study experimentally unchallenged chaos properties, especially in the field of noise filtering.

Remarkably, chaos is weakly sensitive to noise[46,47], implying that a stochastic process will have a negligible effect on the chaotic regime of a system. Inversely, a deterministic signal coupled to a chaotic system could trigger a bifurcation from the chaotic to the periodic regime, hereby amplifying the detection of the deterministic signal. Combining both properties, a chaotic system becomes a noise-free sensor, and numerical simulations demonstrated the detection of signals buried within more than 60dB of noise[11]. This unmatched property lacks experimental demonstration, and M/NEMS are prime tools for such study, being at the frontier between fundamental and applied research.

Through the chaotic regime presented in this paper, combining the noise-free property of chaos with the high sensitivity of M/NEMS is at hand, which not only will create emergent sensors, but also help unravelling the many complex features of this singular chaotic property.

In conclusion, we presented experimentally and numerically a disruptive method to generate a chaotic signal from a nonlinear MEMS structure. The only requirement for the system is to present a Duffing nonlinearity and to be able to perform either amplitude or frequency modulation on the driving signal[30]. We obtained a quantitative comparison between experimental and numerical results, describing the chaotic complexity through the Poincaré sections and the chaotic range through the Lyapunov exponents. Being a model system, M/NEMS enable to explore experimentally properties of chaotic systems, for instance making use of the control on the Lyapunov exponent by tuning the modulation rate. Also, unlike most M/NEMS based chaotic systems, this method does not have any geometrical or material requirement leading to buckling. This freedom enables to implement the chaotic regime in most of the resonant M/NEMS, and we foresee that this could be the first step towards the combination of the high precision features of M/NEMS with the high sensitivity of chaos for sensing applications.

## Methods
### Setup

The voltage source we used is a 33500B Agilent generator, enabling both AM and FM configurations. The measurement of the MEMS device is performed through a HF2TA Zurich Instrument current amplifier with a 10 kΩ load resistance, which output is then probed by a HF2LI Zurich Instrument Lock-In Amplifier. All measured voltages are in root mean square values throughout the paper. Since the bandwidth of the resonator is 50 Hz, the modulation rates of one, two and three times the bandwidth are purposely shifted by 1 Hz to avoid the 50 Hz noise from the electrical lines.

### FM configuration

As suggested by the SI, note 1, the FM configuration requires to demodulate the signal at the modulated frequency of the source. This is easily performed when the generator and the demodulator belong to the same instrument. Otherwise, as in our case, the demodulation frequency has to be synchronized to that of the generator. Without this procedure, both bifurcation diagrams and Poincaré sections would appear extremely noisy, even if the demodulator has a bandwidth much higher than the modulation rate.

Note that the FM modulation is mathematically equivalent to a direct modulation of the resonance frequency $f_0$ instead of the driving frequency $f$, which may be easier to perform depending on the device.

### Numerical simulations

The simulations were performed starting from Eq. 2, with either AM or FM configurations, using the PyDSTool Python library.

### Bifurcation diagrams and Poincaré sections

The datasets processed for the bifurcation diagrams and the Poincaré sections have to be recorded with a very high sampling rate, orders of magnitude higher than the modulation rate. A post-process analysis over under-sampled data (even if the sampling rate is above the spectral bandwidth) results in a poor observation of the $\frac{\delta f}{n}$ bifurcation points and a noisy Poincaré sections. In Fig. 2 and 3, we used a sampling rate of more than two orders of magnitude higher than the modulation rate.

The bifurcation diagrams are obtained by slicing the data at the modulation rate, with any initial phase delay. However, depending on the variable of interest (in our case, the amplitude $R$), some phase delay put more emphasize on the spreading after each bifurcation point. In Fig. 2, we used a convenient initial phase delay between the modulation of the voltage source and the measured signal of 180°, and recorded over 50 cycles of modulation.

This initial phase delay also changes the associated Poincaré sections, displaying more or less complexity. We used a phase delay of 180° in Fig. 3 to stay in line with Fig. 2. However, Poincaré sections require large datasets to reveal their specific signatures, we recorded in this case 1,500 cycles.

### Lyapunov exponents

Each Lyapunov map of Fig. 4 consists in 40x50 pixels, each of them representing a measurement of 200 periods of modulation. The sampling rate was between 30 and 100 times higher than the modulation rate. The Lyapunov exponent is extracted from the dataset using a Wolf algorithm[38] with a relative initial minimal neighbor distance of 3e-3 and a relative final maximal distance of 3e-1.

## Data availability

The data used in this work are available directly from the corresponding authors upon reasonable request.

## Acknowledgments

This project was partially founded by the Federation of Micro and Nanotechnologies (FMNT). The authors thank Gaëtan Debontride for insightful discussions regarding instrumentation techniques.

## Author contributions

M. D. acquired the experimental and numerical data, performed the analysis and developed the analytical model. L. R. did the design of the MEMS. L. F. supervised the TRNG application. S. B. supervised the overall project. M. D. wrote the manuscript with contributions from all the authors.

## Conflict of interest

M. D., L. F. and S. B. declare to have submitted a patent application on the TRNG application under the reference number FR2007766.